# Orthogonalization Properties of Linear Deterministic Polarization Elements

**Sergey N. Savenkov and Yaroslav V. Aulin***

*Quantum RadioPhysics Department, RadioPhysics Faculty, Taras Shevchenko National University of Kyiv, 2 Glushkov Avenue, Kyiv 03022, Ukraine*

*\*Corresponding author: [yaroslav.aulin@gmail.com](mailto:yaroslav.aulin@gmail.com)*

*The conditions under which a linear anisotropic polarization element orthogonalizes several polarization states of input totally polarized light were studied in the paper. The criterion for orthogonalization was obtained in the form of inequality for anisotropy parameters. Orthogonalization properties of polarization elements with the most important anisotropy types were investigated. The parameters under which orthogonalization occurs, and the states that are orthogonalized were found. The loci of these states on the Poincaré sphere were given for sake of illustration in each case.*

## 1. Introduction

Orthogonality plays an important role in polarization optics. The illustrative examples are: horizontal and vertical linearly polarized light; left- and right- circularly polarized. The generalization of the concept of polarization states orthogonality is the orthogonality of two elliptically polarized states with the same form of polarization ellipse, orthogonal major semiaxes and opposite hellicities [1]. On the Poincaré sphere, each totally polarized state is depicted by a point, and two orthogonal polarization states are depicted by diametrically opposite points [1,3] as shown in Fig. 1.

Consider the propagation of totally polarized light beam through linear deterministic anisotropic polarization element. Due to anisotropy of the element the polarization state changes provided it is not eigenstate [1,2]. The case under investigation is when the polarization state of output light beam is orthogonal to the polarization state of corresponding input one.

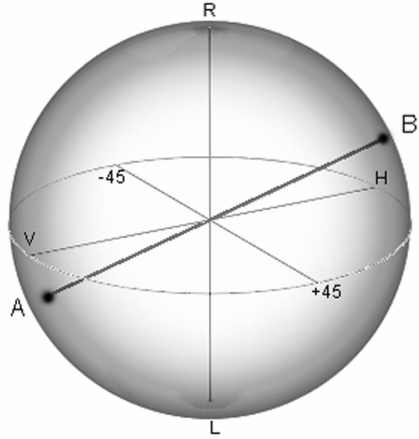

**Fig. 1.** Orthogonal states of polarization A and B on the Poincaré sphere

The goal of the work is to find the circumstances, i.e. the anizotropy parameters of polarization element, under which the orthogonalization occurs at least for one input polarization state, and if the polarization element has orthogonalization properties – the goal is to determine the states that are orthogonalized. Jones matrix calculus [1,2] is the most suitable mathematical description of anisotropic linear deterministic polarization element that interacts with totally polarized light. According to it, medium and field are represented by the Jones matrix **T** and the Jones vector **E** respectively:

$$\mathbf{T} = \begin{bmatrix} T_{11} & T_{12} \\ T_{21} & T_{22} \end{bmatrix}, \quad (1)$$

$$\mathbf{E} = \begin{bmatrix} E_x \\ E_y \end{bmatrix}. \quad (2)$$

In the problem being solved only the information on polarization state of light and its changes after interaction with the polarization element is important, the information on the light wave amplitude is secondary, thus it could be omitted by introducing complex parameter $\chi$ that equals the ratio of field phasors [1-3]:

$$\chi = \frac{E_y}{E_x}. \quad (3)$$

The relation between complex parameters at the input and output, $\chi_{in}$ and $\chi_{out}$ respectively, is represented by bilinear transform:

$$\chi_{out} = \frac{T_{22}\chi_{in} + T_{21}}{T_{12}\chi_{in} + T_{11}}. \tag{4}$$

Two polarization states are orthogonal if the scalar product of corresponding Jones vectors equals zero. The criterion of orthogonality of two states $\chi_{in}$ and $\chi_{out}$ in terms of $\chi$ has the following notation:

$$\chi_{in}\chi_{out}^* = -1. \tag{5}$$

## 2. The inequality for orthogonalization in terms of Jones matrix elements

We are going to obtain the criterion for Jones matrix elements under which orthogonalization for the input state $\chi_{in}$ occurs. Substitution of (4) into (5) yields [3]:

$$T_{11} + T_{12}\chi_{in} + T_{21}\chi_{in}^* + T_{22}|\chi_{in}|^2 = 0. \tag{6}$$

Assuming $T_{22} \neq 0$, we obtain

$$\frac{T_{11}}{T_{22}} + \frac{T_{12}}{T_{22}}\chi_{in} + \frac{T_{21}}{T_{22}}\chi_{in}^* + |\chi_{in}|^2 = 0. \tag{7}$$

Equation (7) is the equation for $\chi_{in}$, where $T_{mn}$ are parameters. In order to solve (7) and find $\chi_{in}$ we substitute

$$\chi_{in} = x + jy, \tag{8}$$

$$T_{mn} = X_{mn} + jY_{mn}. \tag{9}$$

where $X_{mn}, Y_{mn}, x_{mn}, y_{mn}$ are real values.

Thus we have obtained the system of two equations in terms of real variables with real coefficients:

$$\begin{cases} \left(x + \frac{a_1}{2}\right)^2 + \left(y + \frac{b_1}{2}\right)^2 = \frac{a_1^2 + b_1^2}{4} - c_1, \\ a_2 x + b_2 y + c_2 = 0. \end{cases} \tag{10}$$

The following parameters were introduced:

$$a_1 = \frac{X_{22}(X_{12} + X_{21}) + Y_{22}(Y_{12} + Y_{21})}{X_{22}^2 + Y_{22}^2}, \tag{11}$$

$$b_1 = \frac{X_{22}(Y_{21} - Y_{12}) + Y_{22}(X_{12} - X_{21})}{X_{22}^2 + Y_{22}^2}, \tag{12}$$

$$c_1 = \frac{X_{11}X_{22} + Y_{11}Y_{22}}{X_{22}^2 + Y_{22}^2}, \tag{13}$$

$$a_2 = X_{22}(Y_{12} + Y_{21}) - Y_{22}(X_{12} + X_{21}), \tag{14}$$

$$b_2 = X_{22}(X_{12} - X_{21}) + Y_{22}(Y_{12} - Y_{21}), \tag{15}$$

$$c_2 = X_{22}Y_{11} - Y_{22}X_{11}. \tag{16}$$

The analysis of system (10) leads to the conclusion that it has solutions provided the following condition is satisfied:

$$\left(a_1^2 + b_1^2 - 4c_1\right)\left(a_2^2 + b_2^2\right) - \left(a_1 a_2 + b_1 b_2 - 2c_2\right)^2 \geq 0. \tag{17}$$

Due to the analytical simplifications made, the condition whether parameter $R_0^2$ is nonnegative should be verified:

$$R_0^2 = \frac{a_1^2 + b_1^2}{4} - c_1. \tag{18}$$

In terms of real and imaginary parts of Jones matrix elements the inequality (7) has the following form:

$$\begin{aligned}
F_0 &= X_{12}^4 + X_{21}^4 + Y_{12}^4 + Y_{21}^4 + 2\left(X_{12}^2 Y_{12}^2 + X_{21}^2 Y_{21}^2 - X_{21}^2 Y_{12}^2 - X_{12}^2 Y_{21}^2 - X_{12}^2 X_{21}^2 - Y_{12}^2 Y_{21}^2\right) \\
&\quad -4\left(X_{11}^2 Y_{22}^2 + X_{22}^2 Y_{11}^2\right) - 4\left(X_{11}X_{22} + Y_{11}Y_{22}\right)\left(X_{12}^2 + X_{21}^2 + Y_{12}^2 + Y_{21}^2\right) \\
&\quad + 8\left(X_{11}X_{12}X_{21}X_{22} + Y_{11}Y_{12}Y_{21}Y_{22} + X_{12}X_{22}Y_{11}Y_{21} + X_{21}X_{22}Y_{11}Y_{12} \right. \\
&\quad \left. + X_{11}X_{21}Y_{12}Y_{22} + X_{11}X_{12}Y_{21}Y_{22} + X_{11}X_{22}Y_{11}Y_{22} - X_{11}X_{22}Y_{12}Y_{21} - X_{12}X_{21}Y_{11}Y_{22}\right) \geq 0.
\end{aligned} \tag{19}$$

## 3. Generalized decomposition of the Jones matrix. Anisotropy parameters

The Jones matrix depends on eight real parameters, but if the polarization characteristics of the polarization element are of primary value, there exists a possibility to reduce the number of independent parameters to six by omitting isotropic absorption and general phase shift. The simplest way of parameterization of an arbitrary Jones matrix is the generalized Jones matrix decomposition [4,5], according to which the matrix is decomposed into the product of four matrices of principal anisotropy mechanisms: linear and circular amplitude and linear and circular phase anisotropies :

$$\mathbf{T} = [Cir.Ph] \cdot [Lin.Ph] \cdot [Lin.Amp] \cdot [Cir.Amp]. \qquad (20)$$

In (20) the complex factor that denotes isotropic absorption and general phase shift has been omitted.

The Jones matrices of principal anisotropy mechanisms are $[Lin.Amp]$, $[Cir.Amp]$, $[Lin.Ph]$ and $[Cir.Ph]$:

$$[Cir.Ph] = \begin{bmatrix} \cos(\varphi) & \sin(\varphi) \\ -\sin(\varphi) & \cos(\varphi) \end{bmatrix}, \qquad (21)$$

$$[Lin.Ph] = \begin{bmatrix} \cos^2(\alpha) + \sin^2(\alpha)\exp(-j\delta) & \cos(\alpha)\sin(\alpha)(1-\exp(-j\delta)) \\ \cos(\alpha)\sin(\alpha)(1-\exp(-j\delta)) & \sin^2(\alpha) + \cos^2(\alpha)\exp(-j\delta) \end{bmatrix}, \qquad (22)$$

$$[Lin.Amp] = \begin{bmatrix} \cos(\theta)^2 + \sin(\theta)^2 \sqrt{P} & \cos(\theta)\sin(\theta)(1-\sqrt{P}) \\ \cos(\theta)\sin(\theta)(1-\sqrt{P}) & \sin(\theta)^2 + \cos(\theta)^2 \sqrt{P} \end{bmatrix}, \qquad (23)$$

$$[Cir.Amp] = \begin{bmatrix} 1 & -jR \\ jR & 1 \end{bmatrix}. \qquad (24)$$

The parameters $P, \theta, \delta, \alpha, R, \varphi$ are so-called anisotropy parameters and are defined within the intervals [4]: $P \in [0,1]; R \in [-1,1]; \theta \in [0,\pi]; \delta \in [0,2\pi]; \alpha \in [0,\pi]; \varphi \in [0,\pi]$.

The main benefit of using the model (20) is its physical validity, i.e. the medium is described in terms of real physical parameters. Decomposition (20) postulates the equivalence between the polarization element under investigation and the succession of four layers of media with basic anisotropy types and the anisotropy parameters $P, \theta, \delta, \alpha, R, \varphi$. These parameters have the following physical meaning: $R = \dfrac{r_r - r_l}{r_r + r_l}$ – is the value of circular amplitude anisotropy, $P = \dfrac{p_\perp}{p_\parallel}$ – is the value of linear amplitude anisotropy ($r_r$ and $r_l$ are transmission coefficients for right and left circular polarized light and $p_\parallel$ and $p_\perp$ are transmission coefficients for vertical and horizontal linearly polarized). $\theta$ is the maximum transmission axis azimuth, $\delta$ is the value of linear phase anisotropy, $\alpha$ –quick axis azimuth, $\varphi$ –the value of circular phase anisotropy. For further convenience we introduce parameter $\beta = \alpha - \theta$, which is the internal parameter of polarization element. By introducing $\beta$ we exclude the azimuth of polarization element. Finally inequality (19) transforms to

$$F(\varphi, \beta, \delta, P, R) \geq 0. \tag{25}$$

Function $F$ has an explicit form [8]:

$$\begin{aligned}
F =\ & C_0 + C_1 \cos(\delta) + C_2 \cos(2\delta) + C_3 \cos(2\varphi) + C_4 \cos(2\delta)\cos(2\varphi) + \\
& + C_5 \cos^4\left(\frac{\delta}{2}\right)\cos(4\varphi) + C_6 \cos(4\beta - 2\varphi) + C_7 \cos(2\delta)\cos(4\beta - 2\varphi) + \\
& + C_8 \sin^4\left(\frac{\delta}{2}\right)\cos(8\beta - 4\varphi) + C_9 \sin^2(\delta)\cos(4\beta) + C_{10} \sin^2(\delta)\cos(4\beta - 4\varphi) + \\
& + C_{11} \sin(\delta)\sin(2\beta) + C_{12} \sin(2\delta)\sin(2\beta) + C_{13} \cos\left(\frac{\delta}{2}\right)\sin^3\left(\frac{\delta}{2}\right)\sin(6\beta - 2\varphi) + \\
& + C_{14} \sin\left(\frac{\delta}{2}\right)\cos^3\left(\frac{\delta}{2}\right)\sin(2\beta + 2\varphi) + C_{15} \sin(\delta)\sin(2\beta - 2\varphi) + \\
& + C_{16} \sin(2\delta)\sin(2\beta - 2\varphi) + C_{17} \cos(\delta)\sin(2\beta)\sin(2\beta - 2\varphi),
\end{aligned} \tag{26}$$

the coefficients introduced, $C_i$ depend on the values of amplitude anisotropies in decomposition (20):

$$C_i = C_i(P, R), \tag{27}$$

and are as follows:

$$C_0 = (11 + 8R^2 + 11R^4)(1 + P^2) + 2P(134R^2 - 39(1 + R^4)), \tag{28}$$

$$C_1 = -16\sqrt{P}(1 + P)(1 + 6R^2 + R^4), \tag{29}$$

$$C_2 = (1 - 10R^2 + R^4)(1 + P^2) + 2P(18R^2 + 11(1 + R^4)), \tag{30}$$

$$C_3 = 8(1 + \sqrt{P})^2 (1 - \sqrt{P} + P)(R^4 - 1), \tag{31}$$

$$C_4 = 8(1 + \sqrt{P})^2 \sqrt{P}(R^4 - 1), \tag{32}$$

$$C_5 = 4(1 + \sqrt{P})^4 (1 - R^2)^2, \tag{33}$$

$$C_6 = 8(1 - \sqrt{P})^2 (1 + \sqrt{P} + P)(R^4 - 1), \tag{34}$$

$$C_7 = 8(1 - \sqrt{P})^2 \sqrt{P}(R^4 - 1), \tag{35}$$

$$C_8 = 4\left(1-\sqrt{P}\right)^4 \left(1-R^2\right)^2, \tag{36}$$

$$C_9 = 2(1-P)^2 \left(1-10R^2 + R^4\right), \tag{37}$$

$$C_{10} = 2(1-P)^2 \left(1-R^2\right)^2, \tag{38}$$

$$C_{11} = -32R\left(1-P^2\right)\left(1+R^2\right), \tag{39}$$

$$C_{12} = 32R\sqrt{P}(1-P)\left(1+R^2\right), \tag{40}$$

$$C_{13} = 32\left(1-\sqrt{P}\right)^3 \left(1+\sqrt{P}\right)R\left(1-R^2\right), \tag{41}$$

$$C_{14} = 32\left(1-\sqrt{P}\right)\left(1+\sqrt{P}\right)^3 R\left(1-R^2\right), \tag{42}$$

$$C_{15} = 32R\sqrt{P}(1-P)\left(1-R^2\right), \tag{43}$$

$$C_{16} = 8R\left(1-P^2\right)\left(1-R^2\right), \tag{44}$$

$$C_{17} = 16(1-P)^2 \left(R^4-1\right). \tag{45}$$

Thus the problem has been formulated in terms of inequality with five independent parameters. The number of independent parameters was reduced from eight to five: two parameters were excluded due to isotropic attenuation and general phase shift and one parameter was excluded because of the independence of the fact of the orthogonalization properties existance on the azimuth of the polarization element.

From the first sight it may seem that the analysis in terms of anisotropy parameters is rather more complicated than in terms of the Jones matrix elements, but actually it is not true because the approach of anisotropy parameters allows to reduce the number of independent parameters in the problem being solved, these parameters have direct physical interpretation and the formulae are quite possible to deal with using computer facilities.

As the examples of implementation of the results obtained we will study orthogonalization properties for polarization elements with several anisotropy types: circular phase anisotropy; linear phase anisotropy; circular amplitude anisotropy; linear amplitude anisotropy; polarization elements, which correspond first and second Jones equivalence theorems; the polarization element with generalized circular anisotropy.

## 4. Polarization element with circular phase anisotropy

The Jones matrix for a medium with circular anisotropy is $[Cir.Ph]$. For the matrix(21), function (26) gets the following form:

$$F = -1 + \cos(4\varphi). \tag{46}$$

Inequality (25) has only one solution in this case:

$$\varphi = \frac{\pi}{2}. \tag{47}$$

The solution $\varphi = 0$ was omitted because we have $R_0^2 = -1 < 0$ for it.

For (47) the Jones matrix has the following numerical form

$$\begin{bmatrix} 0 & 1 \\ -1 & 0 \end{bmatrix}. \tag{48}$$

Thus equation (7) transforms to

$$\chi_{in} - \chi^*_{in} = 0, \tag{49}$$

i.e. $\text{Im}(\chi_{in}) = 0$, what represents all linearly polarized states. The corresponding points are situated on the equator of the Poincaré sphere as shown in Fig. 2.

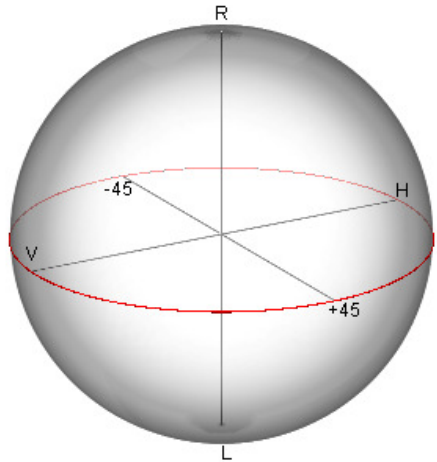

**Fig. 2.** The locus of the input states that are orthonalized on the Poincaré sphere for the polarization element with circular phase anisotropy for $\varphi = \frac{\pi}{2}$.

## 5. Polarization element with linear phase anisotropy

For the polarization element with linear phase anisotropy the Jones matrix is (25). The function (26) is

$$F = -64 + 64\cos(2\delta), \qquad (50)$$

so the inequality (25) transforms to

$$-1 + \cos(2\delta) \geq 0. \qquad (51)$$

Inequality (51) has two solutions: $\delta = 0$ and $\delta = \pi$. If $\delta = 0$, $\mathbf{T}$ is identity matrix and the orthogonalization does not occur, because $R_0^2 = -1 < 0$. If $\delta = \pi$ the orthogonalization occurs because $R_0^2 = 1 > 0$. In this case the Jones matrix has the following view:

$$\begin{bmatrix} \cos(2\alpha) & \sin(2\alpha) \\ \sin(2\alpha) & -\cos(2\alpha) \end{bmatrix}. \qquad (52)$$

For the quick axis azimuth $\alpha = 0$, equation (7) transforms to

$$x^2 + y^2 = 1, \qquad (53)$$

what represents the states with the azimuths $\pm\dfrac{\pi}{4}$. These states are shown on the Poincaré sphere in Fig. 3.

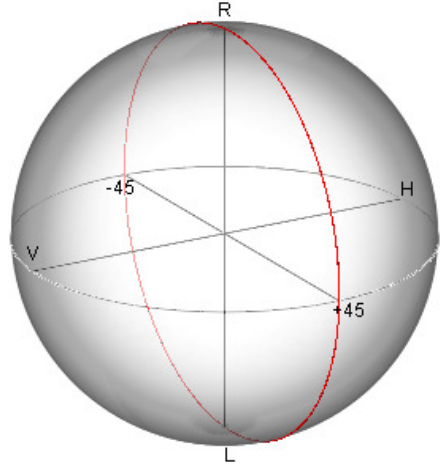

**Fig. 3.** The locus of the input states that are orthonalized on the Poincaré sphere for the polarization element with linear phase anisotropy with $\delta = \pi$ for $\alpha = 0$.

The case $\alpha \neq 0$ gives the curves similar to those in Fig. 3. but rotated along the RL axis by $2\alpha$. This result is illustrated in Fig. 4.

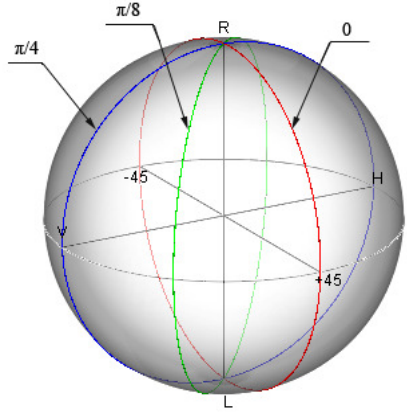

**Fig. 4.** The dependence of the input states that are orthonalized for the polarization element with linear phase anisotropy with $\delta = \pi$ versus $\alpha$.

The rotation of optical system by the angle of $\alpha$ gives the angle of rotation $2\alpha$ around the RL axis on the Poincaré sphere.

## 6. Polarization elements with circular and linear amplitude anisotropies

The Jones matrix for the polarization element with circular amplitude anisotropy is (24), it does not depend on the azimuthal rotation of optical system. The function (26) equals zero for all values of $R$, but $R_0^2 = R^2 - 1$, so the orthogonalization could occur only for $R = \pm 1$. Equation (7) transforms to

$$x^2 + (y \mp 1)^2 = 0 \tag{54}$$

Thus we obtain

$$x = 0;\, y = \pm 1 \tag{55}$$

In the case $R = 1$ we obtain $\chi = j$ – right circular polarization and in the case $R = -1 \text{-} \chi = -j$ – left circular. It could be shown easily that the output intensity is null in both cases.

The case of linear amplitude anisotropy is similar to the case of circular amplitude anisotropy. The orthogonalization properties could be present only in the case of total polarizer ($P = 0$) and the input polarization state that is orthogonalized is linearly polarized in the plane parallel to the absorption axis of polarizer. So, this gives the null intensity at the output.

# 7. Polarization element which corresponds the first Jones equivalence theorem

We will consider the medium with the phase anisotropy in general case. The matrix of this medium according to the first Jones equivalence theorem [6,7] is the product of the matrices of circular and linear phase anisotropies:

$$\mathbf{T} = [Cir.Ph] \cdot [Lin.Ph]. \tag{56}$$

Matrix (56) is the generalized Jones matrix for an arbitrary medium without absorption, the proof of this fact is given in [7]. For example, (6) can be used as the model of polarization properties of a transparent crystal. The inequality (25) transforms to

$$\cos^2\left(\frac{\delta}{2}\right)\cos^2(\varphi) \cdot \left[-6 + 2\cos(\delta) + 2\cos(2\varphi) + \cos(\delta - 2\varphi) + \cos(\delta + 2\varphi)\right] \geq 0. \tag{57}$$

While analyzing (57) we come to the conclusion that only the equality in (57) can be realized. This occurs when the anisotropy parameters have the following values:

a) $\delta = \pi; \forall \varphi.$    b) $\varphi = \dfrac{\pi}{2}; \forall \delta.$

The case ($\delta = 0$; $\varphi = 0$) is omitted because for it we have $R_0^2 = -1$.

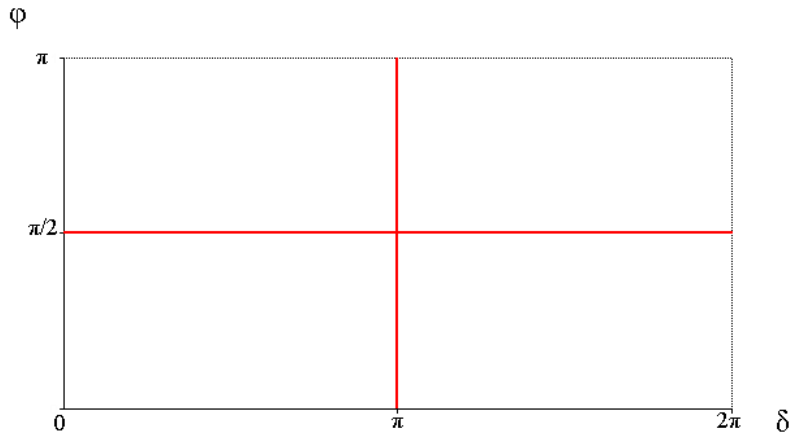

**Fig. 5.** The locus of points on the $(\varphi, \delta)$ plane of system parameters for which the polarization element (56) has the orthonalization properties.

The values of $\varphi$ and $\delta$ for which the orthogonalization occurs can be represented on a $(\varphi, \delta)$ plane of polarization element's parameters.

For (a) the Jones matrix is

$$T = \begin{pmatrix} \cos\gamma & \sin\gamma \\ \sin\gamma & -\cos\gamma \end{pmatrix}. \tag{58}$$

where $\gamma = 2\alpha - \varphi$. We have $a_2 = b_2 = c_2 = 0$. The states that are orthogonalized are situated on the circle on the complex plane:

$$(x - \text{tg}\,\gamma)^2 + y^2 = \frac{1}{\cos^2\gamma}. \tag{59}$$

and if $\cos\gamma = 0$ – on the line $y = 0$.

The representation of these states on the Poincaré sphere is given in Fig. 6.

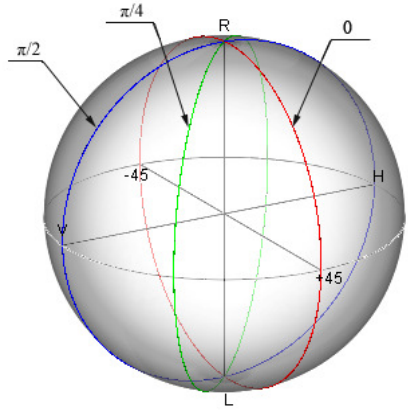

**Fig. 6.** The dependence of the input states that are orthogonalized for the polarization element with the generalized phase anisotropy with $\delta = \pi$ versus $\gamma$.

The states of right circular (R) and left circular (L) polarizations are orthogonalized for arbitrary values of $\gamma$.

For b) like for a) we have $a_2 = b_2 = c_2 = 0$ and the states that are orthogonalized are situated on a circle on the complex plane:

$$(x+\operatorname{ctg}(2\alpha))^2 + \left(y - \frac{1}{\sin(2\alpha)}\operatorname{ctg}\left(\frac{\delta}{2}\right)\right)^2 = \frac{1}{\sin^2(2\alpha)\sin^2\left(\frac{\delta}{2}\right)}. \qquad (60)$$

The cases $\alpha = 0, \frac{\pi}{2}$ and $\delta = 0$ require specific investigation. For $\alpha = 0$ the states that are orthogonalized are located on the line $y = \operatorname{tg}\left(\frac{\delta}{2}\right)x$ if $\delta \neq \pi$ and on the line $x = 0$ if $\delta = \pi$. For $\alpha = \frac{\pi}{2}$ the states that are orthogonalized are located on the line $y = -\operatorname{tg}\left(\frac{\delta}{2}\right)x$ if $\delta \neq \pi$ and on the line $x = 0$ if $\delta = \pi$. For $\delta = 0$ the states that are orthogonalized are located on the line $y = 0$.

The states that are orthogonalized in case b) are shown in Fig. 7 for $\alpha = 0$. For the other values of $\alpha$ we have the same curves, but rotated by the angle of $2\alpha$ around RL axis. As an example the states that are orthogonalized for $\alpha = \frac{\pi}{4}$ are shown in Fig. 8.

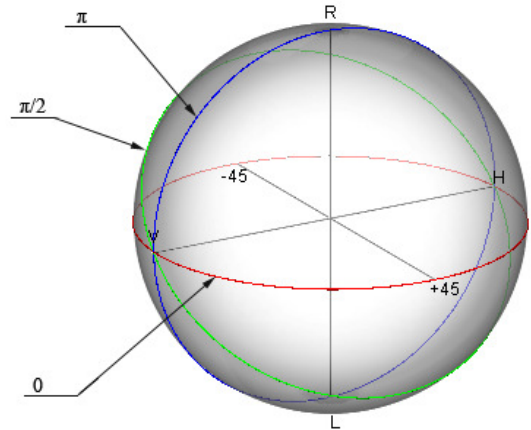

**Fig. 7.** The dependence of the input states that are orthonalized for the polarization element with the generalized phase anisotropy with $\varphi = \frac{\pi}{2}$ versus $\delta$ for $\alpha = 0$.

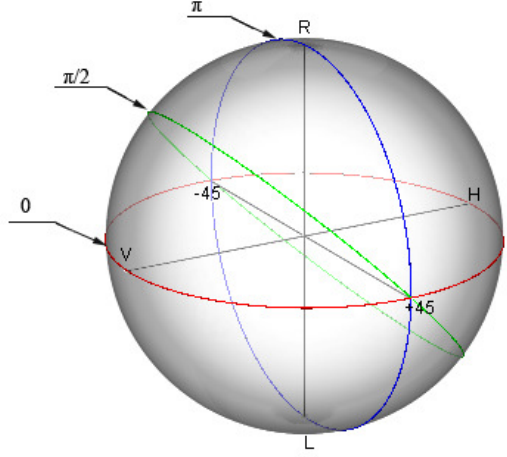

**Fig. 8.** The dependence of the input states that are orthonalized for the polarization element with the generalized phase anisotropy with $\varphi = \dfrac{\pi}{2}$ versus $\delta$ for $\alpha = \dfrac{\pi}{4}$.

## 8. Polarization elements which correspond the second Jones equivalence theorem

Consider the Jones matrix of the following form:

$$\mathbf{T} = [Cir.Ph] \cdot [Lin.Amp]. \tag{61}$$

According to the second Jones equivalence theorem [7], the matrix (6) is the equivalent model of an arbitrary sequence of partial polarizers and optical rotators.

For the Jones matrix (61) we obtain

$$F = -16\left(1+\sqrt{P}\right)^2 \left(-1 + 6\sqrt{P} - P + \left(1+\sqrt{P}\right)^2 \cos(2\varphi)\right) \sin^2(\varphi). \tag{62}$$

The inequality (25) transforms to

$$\left(1 - 6\sqrt{P} + P - \left(1+\sqrt{P}\right)^2 \cos(2\varphi)\right) \sin^2(\varphi) \geq 0. \tag{63}$$

The case $\varphi = 0$ is equivalent to the case of pure linear amplitude anisotropy investigated above. So, for $\varphi \neq 0$ we have

$$\left(1 - 6\sqrt{P} + P - \left(1+\sqrt{P}\right)^2 \cos(2\varphi)\right) \geq 0. \tag{64}$$

For fixed value of $P \in [0,1]$ the inequality (64) is satisfied for the following values of $\varphi$:

$$\frac{1}{2}\arccos\left(\frac{1-6\sqrt{P}+P}{\left(1+\sqrt{P}\right)^2}\right) \leq \varphi \leq \pi - \frac{1}{2}\arccos\left(\frac{1-6\sqrt{P}+P}{\left(1+\sqrt{P}\right)^2}\right). \tag{65}$$

The equality in (65) is realized only at the boundaries of the interval of $P$ values. It's easy to show that $R_0^2 = 0$ for $P = 0$ and $P = 1$, what means that the orthogonalization occurs for these values of anisotropy parameters.

So, we have the region on the $(P,\varphi)$ plane, for anysotropy parameters within what the orthogonalization occurs.

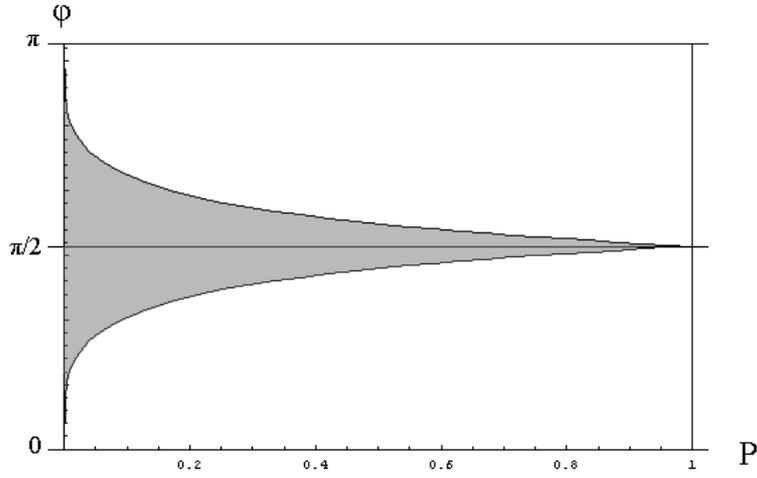

**Fig. 9.** The region on the plane of parameters $(P,\varphi)$ for points within what the polarization element corresponding second Jones equivalence theorem has the orthogonalization properties.

For the polarization element with the Jones matrix (56) there are two input states that are orthogonalized provided condition (65) is satisfied. These states are represented by the following $\chi$ parameters:

$$\chi_{1,2} = \frac{2\left(-1+\sqrt{P}\right)\sin(2\theta-\varphi) \pm \sqrt{2}\sqrt{1-6\sqrt{P}+P-\left(1+\sqrt{P}\right)^2\cos(2\varphi)}}{2\left(\left(-1+\sqrt{P}\right)\cos(2\theta-\varphi)+\left(1+\sqrt{P}\right)\cos(\varphi)\right)}. \tag{66}$$

The values (66) are real, so they represent linearly polarized states.

## 9. Polarization element with circular anisotropy

The Jones matrix for circular anisotropy is

$$\mathbf{T} = [Cir.Ph] \cdot [Cir.Amp]. \tag{67}$$

The inequality (25) transforms to

$$-\left(R^2 - 1\right)^2 \sin^2(2\varphi) \geq 0. \tag{68}$$

It's obvious that only the case of equality could be realized, this occurs for $\varphi = \dfrac{\pi}{2}$ and (or) $R = \pm 1$. In case $R = \pm 1$ the states that are orthogonalized are $\chi = \pm j$ (R and L states respectively), but the output intensity is null.

For $\varphi = \dfrac{\pi}{2}$ the states that are orthogonalized are represented by the circles on the Poincaré sphere:

$$x^2 + \left(y - \frac{1}{R}\right)^2 = \frac{1}{R^2} - 1. \tag{69}$$

The dependence of the states that are orthogonalized versus $R$ is represented in Fig. 10.

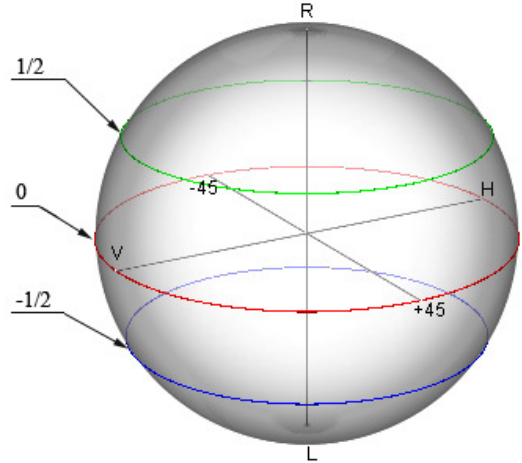

**Fig. 10.** The dependence of the input states that are orthogonalized for the polarization element with circular anisotropy ($\varphi = \dfrac{\pi}{2}$) versus $R$.

## 10. Conclusions

In the work the conditions under which the linear deterministic anisotropic polarization element orthogonalizes several input polarization states of light were obtained. The orthogonalization properties of the elements with basic anisotropy types were studied. It has been shown that the elements with pure linear or circular amplitude anisotropy do not exhibit orthogonalization properties in broad meaning (giving the output orthogonalized polarization with non-zero intensity), while polarization elementts with pure linear or circular phase anisotropy do if $\delta = \pi$ and $\varphi = \dfrac{\pi}{2}$ respectively. The polarization elements that correspond first Jones equivalence theorem have orthogonalization properties if the values of anisotropy parameters are: $\delta = \pi$, or (and) $\varphi = \dfrac{\pi}{2}$. For the polarization elements that correspond the second Jones equivalence theorem there exists a region on the plane of anisotropy parameters $(P, \varphi)$ for points within which the orthogonalization occurs. The states that are orthogonalized in all the cases mentioned were found.